# Concatenated Polar Codes

Mayank Bakshi[1], Sidharth Jaggi[2], Michelle Effros[1]
[1] California Institute of Technology  [2] Chinese University of Hong Kong

*Abstract*—Polar codes have attracted much recent attention as the first codes with low computational complexity that provably achieve optimal rate-regions for a large class of information-theoretic problems. One significant drawback, however, is that for current constructions the probability of error decays sub-exponentially in the block-length (more detailed designs improve the probability of error at the cost of significantly increased computational complexity [1]). In this work we show how the the classical idea of code concatenation – using "short" polar codes as inner codes and a "high-rate" Reed-Solomon code as the outer code – results in substantially improved performance. In particular, code concatenation with a careful choice of parameters boosts the rate of decay of the probability of error to almost exponential in the block-length with essentially no loss in computational complexity. We demonstrate such performance improvements for three sets of information-theoretic problems – a classical point-to-point channel coding problem, a class of multiple-input multiple output channel coding problems, and some network source coding problems.

## I. INTRODUCTION

Polar codes [2] are provably capacity-achieving codes for the Binary Symmetric Channel, with encoding and decoding complexities that scale as $\mathcal{O}(N \log N)$ in the block-length $N$. Polar codes have since demonstrated their versatility – capacity-achieving low-complexity schemes based on polar codes have been demonstrated for a wide variety of source and channel coding problems – for instance, some point-to-point discrete memoryless channels, some rate-distortion problems, the Wyner-Ziv problem and the Gelfand-Pinsker problem [3].

A significant drawback remains. Under the original definition of polar codes their minimum distance can be shown [4] to grow no faster than $-o(\sqrt{N})$ in the block-length $N$. Hence the probability of error decays no faster than $\exp(-o(\sqrt{N}))$ (compared with the $\exp(-\Theta(N))$ probability of error provable for random codes [5]). Low-complexity codes achieving this performance have been constructed [6][1]. Further work to improve the decay-rate of the error probability was partially successful – a sequence of codes have been constructed [1] that in the limit (of a certain implicit parameter denoted $l$) achieves $\exp(-o(N))$ probability of error; however this improvement comes at the expense of significantly increased computational complexity, which scales as $\mathcal{O}(2^l N \log N)$.

In this work we demonstrate that concatenating short polar codes with a high-rate outer Reed-Solomon code significantly improves the rate of decay of the probability of error, with little or no cost in computational complexity. For the point-to-point channel coding problem we use capacity-achieving polar codes of block-length $\Theta(\log^3 N)$ as the inner codes.

[1]The original work [2] only proved that the probability of error decayed as $\mathcal{O}(N^{-1/4})$.

There are three cases of interest. The first case is at one extreme, in which "many" of the inner codes (at least a $\log^{-3/2} N$ fraction) fail resulting in a decoding error with probability $\exp(-\Omega(N/(\log^{-27/8} N)))$ for our concatenated coding scheme. This is the only scenario in which our scheme decodes erroneously.

The second scenario is at the other extreme, in which *none* of the inner codes fail. We show here that if the outer code is a systematic Reed-Solomon code with a rate that approaches 1 asymptotically as a function of $N$, the decoder can quickly verify and decode to the correct output with computational complexity $\mathcal{O}(N(poly(\log N)))$. We show further that the probability of this scenario occurring is $1 - o(1/N)$, hence this is the probabilistically dominant scenario.

The third scenario is the intermediate regime in which at least one (but fewer than a $\log^{-3/2} N$ fraction) of the inner codes fail. Here we show that the Reed-Solomon outer code can correct the errors in the outputs of the inner codes. The complexity of decoding in this scenario $\mathcal{O}(N^2)$ is dominated by the Berlekamp-Massey decoding algorithm for Reed-Solomon codes [7]. However, since this scenario occurs with probability $o(1/N)$, the *average* computational complexity is still dominated by the second scenario.

We then extend these techniques to two other classes of problems. The first class is a general class of multiple-input multiple-output channels, which include as special cases the multiple-access channel and the degraded broadcast channel. The second class is that of network source coding, which includes as a special case the Slepian-Wolf problem. [8] Prior to polar codes no provable low-complexity capacity achieving schemes were known that achieved the optimal rates for these problems – in all cases our codes improve on the rate of decay of the probability of error that polar codes attain, while leaving other parameters of interest essentially unchanged.

Our concatenated code constructions preserve the linearity of the polar codes they are built upon. This is because both the inner polar codes and the outer Reed-Solomon code have a linear encoding structure, albeit over different fields ($\mathbb{F}_2$ and $\mathbb{F}_q$ respectively). However, if we choose the field for the outer code so that $q = 2^r$ for some integer $r$, all linear operations required for encoding over $\mathbb{F}_q$ may be implemented as $r \times r$ matrix operations over $\mathbb{F}_2$. Hence the encoding procedures for both the inner and the outer code may be composed to form a code that is overall a linear code over $\mathbb{F}_2$.

The remainder of the paper is as follows. Section II gives the relevant background of Reed-Solomon codes, polar codes, and concatenated codes respectively. The required notation for this work is summarized in Figure 1. Section III describes our concatenated channel coding scheme, first for the point-to-point channel, and then for a general class of multiple-input multiple-output channels. Section IV describes our corresponding concatenated source coding scheme.

## II. BACKGROUND

**A. Polar codes [2]:** are provably capacity-achieving codes with low encoding and decoding complexity for arbitrary binary-input symmetric discrete memoryless channels. To simplify presentation we focus on binary symmetric channels[2] [8] (BSC($p$)) though many of the results can be generalized to other channels [3]. A crucial component of a polar code is a binary $l \times l$ "base matrix" denoted $G$ (polar encoders and decoders depend on the matrix resulting when $G$ is tensored with itself multiple times). We replicate here some important properties of polar codes relevant for this work.

Let the *inner code rate* of the polar code be denoted by $R_I$. The polar encoder takes as input $R_I n$ bits. The polar encoder outputs a sequence of $n$ bits that are then transmitted over the channel. As the channel is a BSC($p$), the channel flips each bit independently with a probability $p$. The polar decoder then attempts to reconstruct the encoder's $R_I n$ bits.

*Probability of error:* The best known rate of decay of the probability of error of polar codes with increasing block-length $M$ (see [1], [3], [6]) is $(\exp(-o(M^{\beta(l)})))$. Here $\beta(l)$, called the *exponent of the polar code* is a function that is challenging to compute[3], but for instance it is known that $\beta(2) = 0.5$, $\beta(l) \leq 0.5$ for $l \leq 15$, and $\beta(l) \leq 0.6$ for $l \leq 30$.

*Complexity:* Both the encoding and decoding complexity of polar codes scale as $\mathcal{O}(2^l n \log n)$.

*Rate:* While an exact characterization of the speed of convergence of the rate to the Shannon capacity is not known, it is known that polar codes are asymptotically rate-optimal. In this work we denote the (unknown) *rate redundancy* of polar codes by $\delta(n)$.

We note that the rate of decay of the probability of error can be traded off with the computational complexity of polar codes. However, due to the exponential dependence of the computational complexity on the parameter $l$, this cost may be significant for codes that have an exponent close to 1.

*Other rate-optimal channel codes:* There has been much attention on the excellent empirical performance (asymptotically capacity achieving, low encoding and decoding complexity, fast decay of the probability of error) of LDPC codes [9]. However, as of now these results are still not backed up by theoretical guarantees. On the other side, the state of the art in provably good codes are those of Spielman et al. [10], which are asymptotically optimal codes that have provably good performance in terms of computational complexity and probability of error. However, these codes too have their limitations – their computational complexity blows up as the rate of the code approaches capacity.

**B. Reed-Solomon codes [11]:** (henceforth RS codes) are classical error-correcting codes. Let the *outer code rate* $R_O$ of the RS code be any number in $(0, 1)$. The RS encoder takes as input $R_O m$ symbols[4] over a finite field $\mathbb{F}_q$ (here the rate $R_O$, field-size $q$, and the outer code's block-length $m$ are code design parameters to be specified later). The RS encoder outputs a sequence of $m$ symbols over $\mathbb{F}_q$ that are then transmitted over the channel. The encoding complexity of RS-codes is low – clever implementations are equivalent to performing a *Fast Fourier Transform over $\mathbb{F}_q$ (henceforth $\mathbb{F}_q$-FFT)* [12]. But this can be done with $\mathcal{O}(m \log m)$ operations over $\mathbb{F}_q$, or equivalently $\mathcal{O}(m \log m \log q)$ binary operations (for large $q$).

The channel is allowed to arbitrarily corrupt up to $m(1 - R_O)/2$ symbols to any other symbols over $\mathbb{F}_q$.

For all sufficiently large $q$ (in particular if $q \geq m$), the standard RS decoder [7] can reconstruct the source's message with no error (as long as the channel has indeed corrupted fewer than $m(1 - R_O)/2$ symbols). The fastest known RS decoders for errors require $\mathcal{O}(m^2 \log m \log q)$ binary operations (for large $q$) [7].

In this work, we are interested in *systematic* RS codes [7]. A systematic RS code is one in which the first $R_O m$ symbols are in fact the same as the input to the RS encoder, and the remaining $(1 - R_O)m$ *parity-check* symbols correspond to the outputs of a generic RS decoder. These are used in our concatenated code constructions to give highly efficient decoding ($\mathcal{O}(m^2 \log m \log q)$ binary operations) of "high-rate" RS codes when, with high probability (w.h.p.) no errors occur. Details are in Section III.

**C. Code concatenation [13]:** proposed by Forney in 1966, means performing both encoding and decoding in two layers. The source information is first broken up into "many" chunks each of "small" size, and some redundancy is added via an *outer code*. Then each chunk is encoded via a separate *inner code*. The result is the transmitted codeword.

The power of the technique lies in the complementary nature of the two layers. It turns out that the while the probability of error of the inner code decays slowly due to the small block-length of each chunk, its encoding and decoding complexities are also low for the same reason. In contrast, since the outer code has many chunks to code over it can achieve a fast rate of convergence of the probability of error. However, since the outer code is over a larger alphabet, algebraic codes with efficient encoding and decoding techniques (such as RS codes) can be employed. Combining the two layers with the appropriate choice of parameters results in an overall code with low encoding and decoding complexity, and also a reasonably fast decay of probability of error.

In the context of channel coding, we summarize our notation in Figure 1. A detailed description of the corresponding code concatenation scheme is given in Section III.

---

[2] Binary-input binary-output channels that flip each bit input to them according to a Bernoulli($p$) distribution.

[3] Upper and lower bounds on the growth of $\beta(l)$ with $l$ are known [1] – these bounds are again not in closed form and require significant computation.

[4] As is standard we assume here that $m/R_O$ is an integer – if not, choosing a large enough $m$ allows one to choose $R'_O \approx R_O$ resulting in codes with approximately the same behaviour in all the parameters of interest.

| Parameter | Meaning | Our parameter choice |
|---|---|---|
| $n$ | Block-length of inner codes (over $\mathbb{F}_2$) | $\log^3 N$ |
| $R_I$ | Rate of inner codes | $1 - H_b(p) - \delta(\log^3 N)$ |
| $p_i$ | Probability of error of inner codes | $\exp(-\Omega(\log^{9/8} N))$ |
| $q = 2^{R_I n}$ | Field-size of outer code | $2^{R_I \log^3 N}$ |
| $m$ | Block-length of outer code (over $\mathbb{F}_q$) | $N \log^{-3} N$ |
| $R_O$ | Rate of outer code | $1 - 2\log^{-3/2} N$ |
| $N = nm$ | Block-length of overall code (over $\mathbb{F}_2$) | $N$ |
| $M = R_I R_O N$ | Number of source bits | $\left(1 - H_b(p) - \delta(\log^3 N)\right)\left(1 - 2\log^{-3/2} N\right) N$ |
| $P_e$ | Probability of error of overall code | $\exp(-\Omega(N \log^{-27/8} N))$ |

Fig. 1. Summary of notation for the binary symmetric channel

## III. CHANNEL CODING

For ease of exposition we first outline our main ideas for a point-to-point channel. In particular, we focus on the binary symmetric channel BSC($p$), though our results can directly be extended to more general scenarios.

As is well-known, the optimal rate achievable asymptotically for large block-length for a BSC($p$) equals $1 - H_b(p)$, where $H_b(.)$ refers to the *binary entropy function* [8].

### A. Binary symmetric channel

For each $N$ we construct our codes as follows.

*1) Encoder:* Let $n = \log^3 N$ be the block-length of the inner polar codes, $R_I = 1 - H_b(p) - \delta(n) = 1 - H_b(p) - \delta(\log^3 N)$ be their rate, and $p_i = \exp(-\Omega(\log^{3\beta} N))$ be their probability of error. (Here $\beta$ is any value in $(1/3, 1/2)$. The upper bound arises due to the provable rate of decay of the probability of error of polar codes [3], and the lower bound arises due to a technical condition required in (1). For concreteness we set $\beta = 3/8$.) Let $f_I : \mathbb{F}_2^{R_I n} :\to \mathbb{F}_2^n$ and $g_I : \mathbb{F}_2^n :\to \mathbb{F}_2^{R_I n}$ denote respectively the encoder and decoder of an inner polar code.

Correspondingly, let $m = N/n = N/\log^3 N$ be the block-length of the outer systematic RS code, $q = 2^{R_I n} = 2^{R_I \log^3 N}$ be the field-size[5] and $R_O = 1 - 2\log^{-3/2} N$ be its rate (so that the corresponding RS decoder can corrupt up to a fraction $\log^{-3/2} N$ of symbol errors). Let $f_O : \mathbb{F}_q^{R_O m} \to \mathbb{F}_q^m$ and $g_O : \mathbb{F}_{2^n}^m \to \mathbb{F}_{2^n}^{R_O m}$ denote respectively the encoder and (Berlekamp-Massey) decoder for the outer systematic RS code.

Let $M = R_O R_I N$. Define the concatenated code through the encoder function $f : \mathbb{F}_2^M \to \mathbb{F}_2^N$ such that for each *source message* $\mathbf{u}^M \in \{0,1\}^M$,

$$f(\mathbf{u}^M) = (f_I(\mathbf{x}_1), f_I(\mathbf{x}_2), \ldots, f_I(\mathbf{x}_m)),$$

where for each $i$ in $\{1, 2, \ldots, m\}$, $\mathbf{x}_i$ represents the $i$th symbol of the output of the outer systematic RS encoder $f_O(\mathbf{u}^M)$, viewed as a length-$R_I n$ bit vector.

As noted in the introduction, since the inner code is linear over $\mathbb{F}_2$, and the outer code is linear over a field whose size is a power of 2 (and as such may be implemented via matrix operations over $\mathbb{F}_2$) the overall encoding operation is linear.

[5] For this choice of parameters $q = \omega(m)$, as required for RS codes.

*2) Channel:* The channel corrupts each transmitted inner code vector $\mathbf{x}_i$ to $\mathbf{y}_i$, resulting in the output bit-vector $\mathbf{y}^N$.

*3) Decoder:* The concatenated polar code decoder proceeds as follows:

1) It decodes each successive $n$ bit-vector $\mathbf{y}_1, \mathbf{y}_2, \ldots, \mathbf{y}_m$ using the inner polar code decoders $g_I$s to the length-$R_I n$ bit vectors $\hat{\mathbf{x}}_1, \hat{\mathbf{x}}_2, \ldots, \hat{\mathbf{x}}_m = g_I(\mathbf{y}_1), g_I(\mathbf{y}_2), \ldots, g_I(\mathbf{y}_m)$.
2) It passes the first $R_O m$ outputs $\hat{\mathbf{x}}_1, \hat{\mathbf{x}}_2, \ldots, \hat{\mathbf{x}}_{R_O m}$ of the inner code decoders through the systematic RS encoder $f_O$.
3) If $f_O(\hat{\mathbf{x}}_1, \hat{\mathbf{x}}_2, \ldots, \hat{\mathbf{x}}_{R_O m}) = \hat{\mathbf{x}}_1, \hat{\mathbf{x}}_2, \ldots, \hat{\mathbf{x}}_m$ then it declares $\hat{\mathbf{x}}_1, \hat{\mathbf{y}}_2, \ldots, \hat{\mathbf{x}}_{R_O m}$ as the decoded message (denoted by $\bar{\mathbf{x}}^M$) and terminates.
4) Otherwise it passes $\hat{\mathbf{y}}_1, \hat{\mathbf{y}}_2, \ldots, \hat{\mathbf{y}}_m$ through the outer decoder $g_O$ (a standard RS Berlekamp-Massey decoder), declares the length-$M$ bit-vector corresponding to the output $\mathbf{x}_1^*, \mathbf{x}_2^*, \ldots, \mathbf{x}_{R_O m}^* = g_O(\hat{\mathbf{x}}_1, \hat{\mathbf{x}}_2, \ldots, \hat{\mathbf{x}}_m)$ as the decoded message (denoted by $\bar{\mathbf{x}}^M$), and terminates.

The rationale for this decoding algorithm is as follows. Step 1 uses the $m$ inner codes to attempt to correct the errors in each symbol of the outer code. If each resulting symbol is indeed error-free, then since the outer code is a systematic RS code, re-encoding the first $R_O m$ symbols (Step 2) should result in the observed decoder output (Step 3). On the other hand, if the inner codes do not succeed in correcting all symbols for the outer code, but there fewer than $(1-R_O)m/2 = N \log^{-9/2} N$ errors in these $m = N \log^{-3} N$ symbols, then the RS outer decoder succeeds in corrected all the outer code symbols (Step 4). Hence an error occurs only if there are $N \log^{-9/2} N$ or more errors in the outer code. The probability of this event can be bounded from above by the following theorem.

**Theorem 1.** *For the concatenated polar code defined above, for all $N$ large enough, $P_e < \exp(-\Theta(N/\log^{27/8} N))$.*

*Proof:* By the polar code exponent of [6], our specific choice of $\beta = 3/8$ in the concatenated polar encoder, and for $n$ large enough, the probability that any specific inner code fails is at most

$$\Pr(g_I(\mathbf{x}_i) \neq \mathbf{x}_i) < \exp(-n^\beta) = \exp(-\log^{9/8} N).$$

As noted above our code construction fails only if $N\log^{-9/2} N$ or more of the $m = N\log^{-3} N$ inner codes fail. Hence the probability of error

$$P_e \leq \binom{N\log^{-3} N}{N\log^{-9/2} N}\left(\exp\left(-\log^{9/8} N\right)\right)^{N\log^{-9/2} N}$$

$$\leq \exp\left(\frac{N}{\log^3 N}H_b\left(\frac{1}{\log^{3/2} N}\right)\right)\left(\exp\left(-\frac{N}{\log^{27/8} N}\right)\right),$$

where the second inequality is due to Stirling's approximation [8]. Next we note that for $\lim_{\epsilon \to 0} H_b(\epsilon)/\epsilon^\alpha = 0$ for every $\alpha \in [0, 1)$. In particular, by choosing $\alpha = 1/2$, we obtain,

$$P_e \leq \exp\left(N(\log^{-3} N)\left(\log^{-3/4} N\right) - \left(N\log^{-27/8} N\right)\right)$$
$$< \exp(\Theta(N\log^{-27/8} N)),$$

for large enough $N$. Finally, we see that this construction is capacity achieving since the inner codes and outer code are constructed at rates asymptotically (in $N$) approaching channel capacity and 1 respectively. ∎

Notice here that with the above choice of parameters $n$ and $m$, the expected number of errors in the received codeword for the outer codeword approaches zero. Therefore, with a high probability, we receive the transmitted codeword without any error. We exploit this fact in showing that the average complexity of the decoding algorithm is dominated by the complexity of the verification step. This is characterized in the following result.

**Theorem 2.** *With the choice of parameters as in Theorem 1, the above algorithm runs in $\mathcal{O}(N \log N)$ time.*

*Proof:* Our inner decoder decodes each of the $\Theta(N/\log^3 N)$ inner codes using the standard polar code decoder. By the standard polar code *successive cancellation* decoding procedure [2], the computational complexity of decoding each inner code is $\mathcal{O}(\log^3 N \log(\log^3 N))$, which equals $\mathcal{O}(\log^3 N \log \log N))$. Since there are $\Theta(N/\log^3 N)$ inner codes, the overall decoding complexity of the inner code decoders is $\mathcal{O}(N \log \log N))$. This is dominated by the next decoding step, and hence we neglect this.

We note that for our code construction the average decoding complexity of a systematic RS code can be reduced so as to approximately equal its encoding complexity ($\mathcal{O}(m \log m \log q)$ binary operations), as follows.

Recall our outer decoder follows the following procedure. It first performs the encoding operation again on the first $R_o m$ symbols and compares the output of this second encoding process with the observed $m$ symbols. If these two sequences are indeed the same the decoder outputs the first $R_o m$ symbols as the decoded output and terminates. If not the decoder then defaults to standard RS decoding [7] to output $R_o m$ symbols as the decoded output and terminates.

Let $P_1$ denote the probability that at least one subblock has been decoded erroneously by the polar decoder. Since $P_1 < m \exp(-n^{3/8})$ for our choice of $\beta = 3/8$, $P_1$ decays as

$$P_1 < \exp(\mathcal{O}((\log N)^{9/8}))(\log^{-3} N) = o(1/m). \quad (1)$$

We now consider the complexity of this decoder for the three scenarios for the outer code.

1) *The channel for the outer code corrupts no symbols, with probability at least $1 - o(1/m)$:* In this case the decoder terminates after the first step, having performed $\mathcal{O}(m \log m \log q)$ elementary operations.
2) *The channel for the outer code corrupts fewer than $m(1 - R_o)/2$ symbols, with probability at most $1 - o(1/m)$:* In this case the decoder terminates after the second step, having performed an additional $\mathcal{O}(m^2 \log m \log q)$ elementary operations.
3) *The channel for the outer code corrupts $m(1 - R_o)/2$ or more symbols, with probability at most $\exp(-\Omega(-N \log^{27/8} N))$:* This case is an alternative for the previous scenario, and also terminates after an additional $\mathcal{O}(m^2 \log m \log q)$ elementary operations.

Thus the overall average computational complexity is $\mathcal{O}(m \log m \log q)$. Recalling that for our choice of parameters $m = \Theta(N/\log^3 N)$ and $q = \exp(\Theta(\log^3 N))$, this gives an overall complexity of $\mathcal{O}(N \log N)$. ∎

Observe that concatenation essentially operates on the message bits and is independent of the channel under consideration. Therefore, if each transmitter has more than one kind of messages, like in the case of a multiple-input multiple-output channel, the operation outlined above can be repeated for each message separately. Likewise, at each decoder, the above decoding rule may be applied for each message after decoding the inner code. Since polar codes have been shown to be optimal for certain multiuser channels [3], and similar results on the rate of polarization continue to hold, we have the following corollary:

**Corollary 1** (Multiple Access Channel and Degraded Broadcast Channel)**.** *For the two user multiple access channel $p_{Y|X_1X_2}(\cdot|\cdot)$ and the degraded broadcast channel $p_{X_1X_2|Y}$, the code constructed by concatenating a polar code of block length $\log^3 N$ with a RS code of block length $N\log^{-3} N$ has an error probability that decays as $\exp(-\Omega(N \log^{-27/8} N))$ and can be encoded and decoded in $\mathcal{O}(N \log N)$.*

## IV. CODE DESIGN FOR NETWORK SOURCE CODING

The results in previous sections demonstrate that concatenation is helpful in reducing the error probability for channel coding. In this section, we show that the error probability for network source coding may be similarly reduced via concatenation.

A network $\mathcal{N} = (V, E)$ is a directed graph. Denote by $S \subseteq V$ and $T \subseteq V$, respectively, the sets of source and sink nodes. Each source node $s \in S$ observes a source random process $U(s)$, and each sink node $t \in T$ demands a subset $W(t) = (U(s) : s \in S_t)$ of these for some $S_t \subseteq S$. The random process $\{(U_i(s) : s \in S)\}_{i=1}^\infty$ is drawn i.i.d. from a known probability mass function $P(\cdot)$. In our discussion, we consider only binary sources. It can be seen that the technique outlined here extend to non-binary sources as well.

For a collection of rates $(R_e \geq 0 : e \in E)$, an $(n, (2^{nR_e})_{e \in E})$ network source code $(F^{(n)}, G^{(n)})$ comprises of encoder maps $F^{(n)} = (f_e^{(n)} : e \in E)$ and decoder maps $G^{(n)} = (g_t^{(n)} : t \in T)$ with

$$f_{(v,v')}^{(n)} : \mathbb{F}_2^n \to \mathbb{F}_2^{nR_{(v,v')}} \forall v \in S$$
$$f_{(v,v')}^{(n)} : \prod_{e \in \Gamma_i(v)} \mathbb{F}_2^{nR_e} \to \mathbb{F}_2^{nR_{(v,v')}} \forall (v, v') \in \Gamma_o(V \setminus S)$$
$$g_t^{(n)} : \prod_{e \in \Gamma_i(t)} \mathbb{F}_2^{nR_e} \to \prod_{s \in S_t} \mathbb{F}_2^n \quad \forall t \in T.$$

The probability of error for the above code is defined as

$$P_e \triangleq \Pr(g_t^{(n)}(f_e : e \in \Gamma_i(t)) \neq W_1(t), \ldots, W_n(t) \text{ for some } t).$$

*A. Concatenated Code Construction*

As in the previous constructions, fix $n = \log^3 M$ and $m = M \log^{-3} M$ to be the input block-lengths for the inner and outer codes. The outer code is chosen to be a systematic RS code. The concatenated code construction for this case is similar to that for channel coding, except for a few differences. For an input block length $M = nm$, the code $(\widetilde{F}^{(M)}, \widetilde{G}^{(M)})$ is constructed as follows.

*1) Encoder:* Let $(F^{(n)}, G^{(n)})$ be a code that operates at a rate $(R_e : e \in E)$ and . Let $R_O = (1 - 2/n^{4\beta/3})$. Let $m = M/(nR_O)$ be the codelength for the outer systematic RS code. Denote the outer code as $f_o : \mathbb{F}_{2^n}^{mR_O} \to \mathbb{F}_{2^n}^m$. Let $g_o$ be the corresponding decoder. Note that since $f_o$ is chosen to be a systematic code,

$$f_o(\mathbf{u}^M(s)) = \mathbf{u}^M(s) * \widetilde{f}_o(\mathbf{u}^M(s))$$

for some function $\widetilde{f}_o$. The encoding is performed in two steps:
  a) Apply the code $(F^{(n)}, G^{(n)})$ to each sub-block $\mathbf{u}_{(i-1)n+1}^{in}(S)$ for $i = 1, \ldots, m$.
  b) For each $s \in S$, multicast $\widetilde{f}_o(\mathbf{u}^{MR_O}(s))$ to each sink $t$ such that $s \in S_t$. The excess rate added on each edge as a result of this transmission is upper bounded by $|S|(1 - R_O)/2$.

*2) Decoder:* Since each sink receives the redundant bits for the outer code for every desired source, as well as the codewords from the inner code, decoding can be performed in the following two steps:
  a) Apply the decoder functions $g_t^{(n)}$ to each sub-block separately to obtain reconstructions $\widetilde{\mathbf{w}}^M(t) = (\widetilde{\mathbf{u}}^M(s,t) : s \in S_t)$.
  b) Output the reconstruction $\widehat{\mathbf{u}}^M(s,t) = g_o^s(\widetilde{\mathbf{u}}^M(s,t), \widetilde{f}_o^s(\mathbf{u}^M(s)))$.

The above construction technique is independent of the choice of the inner code. For specific networks such as Slepian-Wolf network and Ahlswede-Körner network etc, polar codes are shown to be optimal [3]. Further, the error probability is shown to vanish as $\exp(-n^\beta)$ for $\beta < 1/2$. Since the length of the inner code and the outer code, and the rate of the outer code have been chosen in the same way as the concatenated channel code construction, as a corollary of Theorems 1 and 2, we have the following result:

**Corollary 2.** *Let $\mathcal{N}$ be a network such that there exists a sequence of polar codes $\{(F^{(n)}, G^{(n)})\}_{n=1}^\infty$ that are optimal. Then, the sequence of concatenated codes outlined above are also optimal. Further, the error probability decays as $\exp(-\Omega(M \log^{-27/8} M))$ and the encoding, and decoding can be performed in $\mathcal{O}(M \log M)$ time.*

## V. CONCLUSION

In this work we examine the question of the tradeoff between the computational complexity of polar codes and the rate of decay of their probability of error. We demonstrate that using the well-studied technique of concatenation, the rate of decay of the probability of error can be boosted to essentially optimal performance. The question of the corresponding speed of convergence of the code rates to the optimal rate-region is still an interesting open question, as it is for the original formulation of polar codes.

## VI. ACKNOWLEDGEMENTS

The authors gratefully thank Emre Telatar, Satish Korada, and Rüdiger Urbanke for introducing us to the elegant work on polar codes and patiently explaining vexing details.